\documentclass[conference]{IEEEtran}
\IEEEoverridecommandlockouts
\usepackage{cite}
\usepackage{amsmath,amssymb,amsfonts}
\usepackage{algorithmic}
\usepackage{graphicx}
\usepackage{textcomp}
\usepackage{xcolor}
\usepackage{booktabs,ragged2e}
\usepackage[flushleft]{threeparttable}
\usepackage[hyphens]{url}

\def\BibTeX{{\rm B\kern-.05em{\sc i\kern-.025em b}\kern-.08em
    T\kern-.1667em\lower.7ex\hbox{E}\kern-.125emX}}

\makeatletter
\newcommand{\linebreakand}{%
  \end{@IEEEauthorhalign}
  \hfill\mbox{}\par
  \mbox{}\hfill\begin{@IEEEauthorhalign}
}
\makeatother

\IEEEoverridecommandlockouts
\IEEEpubid{\makebox[\columnwidth]{979-8-3315-5925-0/26/\$31.00~\copyright~2026 IEEE \hfill} \hspace{\columnsep}\makebox[\columnwidth]{ }}

\begin{document}

\title{Exploring Human-in-the-Loop Themes in AI Application Development:
An Empirical Thematic Analysis\\
\thanks{*Corresponding author}}
\IEEEpubidadjcol

\author{
\IEEEauthorblockN{1\textsuperscript{st} Parm Suksakul}
\IEEEauthorblockA{
\textit{International School of Engineering}\\
\textit{Faculty of Engineering} \\
\textit{Chulalongkorn University}\\
Bangkok, Thailand \\
6538148321@student.chula.ac.th}
\and
\IEEEauthorblockN{2\textsuperscript{nd} Nathan Kittichaikoonkij}
\IEEEauthorblockA{
\textit{International School of Engineering}\\
\textit{Faculty of Engineering} \\
\textit{Chulalongkorn University}\\
Bangkok, Thailand \\
6538047621@student.chula.ac.th}
\and
\IEEEauthorblockN{3\textsuperscript{rd} Nakhin Polthai}
\IEEEauthorblockA{
\textit{International School of Engineering}\\
\textit{Faculty of Engineering} \\
\textit{Chulalongkorn University}\\
Bangkok, Thailand \\
6538055621@student.chula.ac.th}
\linebreakand
\IEEEauthorblockN{4\textsuperscript{th} Aung Pyae*}
\IEEEauthorblockA{
\textit{International School of Engineering}\\
\textit{Faculty of Engineering} \\
\textit{Chulalongkorn University}\\
Bangkok, Thailand \\
aung.p@chula.ac.th}
}

\maketitle

\begin{abstract}
Developing and deploying AI applications in organizations is challenging when human decision authority and oversight are underspecified across the system lifecycle. Although Human-in-the-Loop (HITL) and Human-Centered AI (HCAI) principles are widely acknowledged, operational guidance for structuring roles, checkpoints, and feedback mechanisms remains fragmented. We report a multi-source qualitative study: a retrospective diary study of a customer-support chatbot and semi-structured interviews with eight AI experts from academia and industry. Through five-cycle thematic analysis of 1,435 codewords, we derive four themes: AI Governance and Human Authority, Human-in-the-Loop Iterative Refinement, AI System Lifecycle and Operational Constraints, and Human–AI Team Collaboration and Coordination. These themes provide empirical inputs for subsequent HITL framework design and validation.
\end{abstract}

\section{Introduction}
Artificial intelligence (AI) application failures in organizations are often framed as technical problems, manifesting as incorrect predictions, unstable performance, or unexpected model behavior \cite{Li_Vorvoreanu_Debellis_Amershi_2023a, NIPS-2015-hidden-technical-debt, Salwei_Carayon_2022}. In this paper, we define an AI application as a production-oriented socio-technical software system that embeds one or more AI components to support or automate tasks within an organizational workflow. Accordingly, when we refer to AI development, we mean the development and lifecycle management of such AI applications rather than foundational model research. While these failures are typically diagnosed at the technical level, we argue that they frequently reflect a deeper socio-technical breakdown \cite{Salwei_Carayon_2022, Dobbe_Wolters_2024, 10.1145/3287560.3287598}, namely the absence of explicit, lifecycle-wide human oversight governing how model outputs are interpreted, challenged, and constrained in operational settings.

This core challenge is amplified by the inherent properties of machine learning (ML) systems, a subfield of AI that serves as a representative example in many organizational deployments. ML components are often embedded within complex software and business processes, where data-driven behavior erodes traditional abstraction boundaries and creates a tight coupling between models, pipelines, and human decisions\cite{NIPS-2015-hidden-technical-debt, breck2017mltestscore, google2023mlops}. As a result, reliability cannot be treated as a static model property established at deployment, but instead emerges across the system lifecycle through continuous monitoring, adaptation, and human oversight \cite{NIPS-2015-hidden-technical-debt, breck2017mltestscore, nistairmf}. Prior work characterizes such systems as “fast to deploy but hard to maintain,” with hidden technical debt accumulating through opaque dependencies and feedback loops \cite{NIPS-2015-hidden-technical-debt, breck2017mltestscore}. This reality shifts the focus from pre-deployment validation alone to the need for continuous, lifecycle-wide governance.

Human-in-the-Loop (HITL) approaches have been proposed as a response to these challenges by positioning human input as an integral component of AI system operation\cite{andersen2023designpatternsmachinelearning, Amershi_Cakmak_Knox_Kulesza_2014, WU2022364, Geissler_Krupp_Banwari_Habusch_Zhou_Lukowicz_Karolus_2025, Retzlaff_Das_Wayllace_Mousavi_Afshari_Yang_Saranti_Angerschmid_Taylor_Holzinger_2024, kandikatla2025aihumanoversightriskbased}. In practice, HITL is frequently instantiated through concrete patterns such as uncertainty-based escalation (routing low-confidence cases to human review)\cite{andersen-maalej-2022-efficient}, human-initialized or model-suggested labeling (e.g., active learning that prioritizes informative instances)\cite{settles.tr09}, and user correction and feedback loops (e.g., accept/reject controls that create traceable supervision signals for refinement\cite{9718470}) \cite{andersen2023designpatternsmachinelearning}. Prior work emphasizes functional allocation, reserving human effort for cases requiring contextual judgment or ethical deliberation \cite{Amershi2019GuidelinesFH}. However, despite effectiveness in narrow settings (e.g., labeling and post-training alignment), HITL remains insufficiently operationalized for organizational AI development, particularly in specifying lifecycle-wide roles, decision authority, and permissible interventions under real-world resource and coordination constraints. 

Complementarily, a study on human-centered AI (HCAI) argues that fairness, accountability, and transparency cannot be achieved through technical mechanisms alone, but require pan-disciplinary collaboration among technologists, socio-technical system design involving collaboration among AI engineers, decision-makers, and affected stakeholders \cite{human-center-ai}. Within this perspective, explainability is positioned not merely as a usability feature, but as a governance mechanism that preserves meaningful human decision authority by enabling traceability of model behavior and evidentiary reasoning \cite{doshivelez2017rigorousscienceinterpretablemachine}. However, while HCAI articulates important normative principles, it provides limited prescriptive guidance on how human roles, responsibilities, and intervention points should be structured operationally within AI development pipelines \cite{Madaio_Stark_WortmanVaughan_Wallach_2020}.

This paper addresses this gap by establishing the empirical foundation needed for operational guidance on human oversight. We report a multi-source qualitative study to answer the question: \textit{What are the key socio-technical themes that characterize the necessary human oversight within AI application development?} We analyze evidence from a retrospective diary study of a customer-support chatbot case study and semi-structured interviews with AI practitioners. Through this thematic analysis, we identify and synthesize recurrent patterns into a set of foundational themes. We define themes as empirically derived, recurring socio-technical patterns that characterize essential structures, relationships, or challenges in human oversight. These themes constitute the primary contribution of this work as preliminary, empirically grounded inputs, subject to further validation and refinement, intended to inform the subsequent design of a structured HITL framework for AI application development.

\section{BACKGROUND}
HCAI frames AI applications as socio-technical systems whose reliability, safety, and organizational value emerge from the interaction of models, data, software infrastructure, and human judgment \cite{shneiderman2020humancenteredartificialintelligencereliable, hermanthomas, Schmager14092025, xu2025hcaimethodologicalframeworkhcaimf}. Within this perspective, HITL extends beyond annotation strategies to a lifecycle design principle that assigns humans explicit responsibility for oversight, interpretation, escalation, and override decisions when system behavior becomes uncertain in real deployment contexts. In this paper, the AI system lifecycle refers to the end-to-end progression of an AI application, including problem formulation, system design, development, deployment, monitoring, maintenance, and subsequent iteration. We use the term Human-in-the-Loop to refer to the systematic integration of human judgment into AI application development and operation, where humans are assigned explicit roles, decision authority, and intervention points across this lifecycle to oversee, challenge, and constrain automated behavior\cite{Meng2023Data, GRONSUND2020101614, NIPS-2015-hidden-technical-debt, WU2022364}.This paper adopts HCAI as the umbrella viewpoint and positions HITL as a concrete operational mechanism for embedding accountable human decision authority into business AI applications, specifically customer-support and analytical chatbots, where outputs are routinely consumed by domain stakeholders and can materially influence organizational decisions.

A substantial portion of HITL research demonstrates the value of targeted human feedback to improve model performance under realistic deployment constraints \cite{wang2018humaninthelooppersonreidentification, christiano2023deepreinforcementlearninghuman, Amershi_Cakmak_Knox_Kulesza_2014}. For example, “Human-In-The-Loop Person Re-Identification” introduces a semi-automated scheme designed to optimize deployment performance given a small number of human verification interactions, emphasizing reduced labeling burden and interactive correction during use rather than relying on large offline labeled datasets \cite{wang2018humaninthelooppersonreidentification}. This line of work is valuable in establishing that human feedback can be strategically allocated to high-uncertainty cases and can yield immediate operational benefit in deployment settings \cite{wang2018humaninthelooppersonreidentification}. However, these studies typically operationalize HITL as an algorithmic feedback loop (verification, correction, incremental learning) without prescribing how professional development teams should define role responsibilities, intervention thresholds, auditability requirements, or organizational escalation pathways across the broader AI lifecycle.

Complementary software engineering research begins to bridge this gap by offering reusable design patterns for ML-based systems with Human-in-the-Loop (HITL) mechanisms. Prior work proposes a catalog of patterns intended to address engineering and deployment challenges and to guide developers in selecting and implementing appropriate HITL solutions across training and deployment contexts \cite{andersen2023designpatternsmachinelearning}. This work explicitly acknowledges that integrating human intelligence into automated decision-making requires additional operational guidance beyond model-centric methods. However, such design pattern catalogs generally do not resolve lifecycle-wide governance questions, including how decision authority is assigned at each stage, how cross-functional stakeholders arbitrate trade-offs, and how human interventions are recorded, reviewed, and institutionalized as part of accountable organizational practice.

Governance frameworks such as the NIST AI Risk Management Framework (AI RMF 1.0) provide an important lifecycle-oriented lens for defining trustworthy outcomes and risk management functions \cite{nistairmf}. NIST organizes AI risk management around four high-level functions—GOVERN, MAP, MEASURE, and MANAGE—emphasizing that governance is cross-cutting and that risk management should be continuous across lifecycle dimensions \cite{nistairmf}. The AI RMF also incorporates feedback channels and consultation with domain experts and other relevant actors as part of measurement and validation in deployment contexts, including processes that enable reporting problems and seeking redress or appeal of system outcomes \cite{nistairmf}. Importantly, NIST clarifies that the listed actions are not a prescriptive checklist and are not necessarily sequenced, reinforcing that the framework is designed to be adaptable rather than procedural \cite{nistairmf}. This adaptability is a strength for broad adoption, but it leaves a persistent operational gap for practitioner teams building business AI applications: the AI RMF specifies the types of outcomes and risk considerations that should be addressed, yet it does not provide concrete, pipeline-level mechanisms for role assignment, decision checkpoints, escalation rules, or implementation artifacts that encode accountable human decision authority in day-to-day development and deployment workflows.

MLOps frameworks provide the engineering foundation for reliable ML delivery and monitoring\cite{google2023mlops,breck2017mltestscore, Kreuzberger_Kuehl_Hirschl_2023}. Google’s Practitioner’s Guide defines MLOps as “standardized processes and technology capabilities for building, deploying, and operationalizing ML systems rapidly and reliably.” emphasizing scaling and automation in production pipelines \cite{google2023mlops}. It highlights continuous training and monitoring practices, alongside artifact tracking and model management, and acknowledges governance needs such as review-and-approval steps and lineage visibility for accountability \cite{google2023mlops}.

Nevertheless, typical MLOps framings position humans primarily as pipeline operators, reviewers, or approvers \cite{google2023mlops,breck2017mltestscore} , but do not systematically specify decision-authority boundaries (e.g., who may override outputs and under what evidentiary conditions), human-feedback interfaces (e.g., how domain experts submit structured corrections), or handoff protocols between technical teams and business stakeholders in AI applications where model outputs shape operational decisions. These omissions are often more consequential in large language model (LLM)–enabled applications, such as customer-support and analytical conversational systems, because failures can be difficult to detect (e.g., hallucinated content\cite{Ji_2023,openai2024gpt4technicalreport, weidinger2021ethicalsocialrisksharm}, policy or domain misalignment\cite{QU2025249}, and underspecified queries\cite{yang2025promptsdontsayunderstanding, Yan_Li_Jiang_Zhao_Guan_Kuo_Wang_2025}), and safe operation frequently depends on human interpretation, escalation, and override decisions during deployment \cite{zou2025llmbasedhumanagentcollaborationinteraction}.

Across HITL algorithmic studies, governance frameworks, and MLOps engineering practices, the literature provides strong components, yet integrated support for end-to-end AI application development remains limited–particularly for specifying human roles, decision authority, and implementable checkpoints that connect governance intent to everyday development and stakeholder workflows. To address this gap, we derive an empirically grounded set of HITL themes from two sources: a diary-based retrospective case analysis of chatbot development in a business setting and semi-structured interviews with AI experts spanning multiple application domains. Each source is independently coded and integrated through thematic analysis to produce a consolidated theme set; framework operationalization and validation are designated as future work.

\section{METHODOLOGY}
\subsection{Study Design and Methodological Rationale}
This research employs a qualitative, multi-source study design integrating two evidence streams: (1) a retrospective diary-based case study documenting the development of an enterprise AI customer-support chatbot, and (2) a semi-structured interview study with AI experts. The objective is to empirically characterize HITL practices as they manifest in real-world AI development, rather than to propose or validate a prescriptive framework.

The diary study captures situated development practices, including decision-making, coordination, and responses to system failures. The interview study complements this perspective by eliciting expert expectations and constraints related to governance, scalability, and accountability in enterprise AI systems. Together, these sources enable triangulation between observed practice and external professional perspectives.

Data from both sources were first coded independently and subsequently consolidated through an iterative thematic synthesis process. The construction and validation of a formal HITL framework based on these themes is reserved for future work.

\subsection{Data Source 1: Retrospective Diary Study and Case Description}
This study draws on a retrospective diary-based case study of a production-oriented AI chatbot developed and deployed within a software enterprise. The system aimed to reduce response latency and improve procedural troubleshooting accuracy through an interactive workflow implemented on the company’s messaging platform. The primary research objective of this case study was the systematic observation, documentation, and formalization of human agency throughout the life-cycle of the AI application.

\begin{figure}
    \centering
    \includegraphics[width=\linewidth]{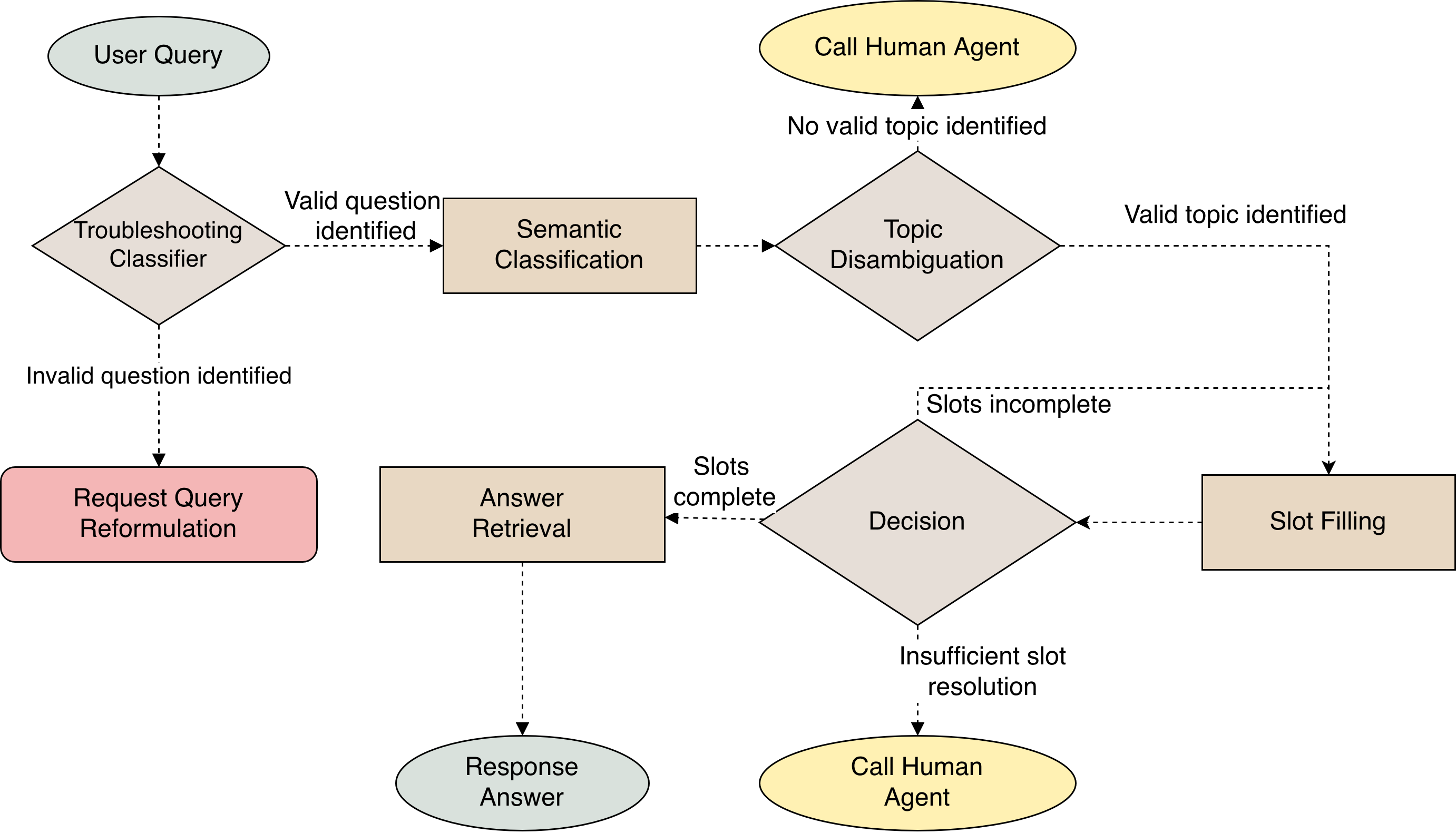}
    \caption{Overview of the  customer support chatbot workflow–business AI application case study}
    \label{fig:LLM_RAG_workflow}
\end{figure}

The project unfolded in a multi-stakeholder socio-technical setting. A central AI engineer translated business needs into system specifications and coordinated iterative alignment among management (strategic priorities), support staff (domain expertise), and software engineers (deployment constraints). An initial proof of concept implemented a Retrieval-Augmented Generation (RAG) prototype using bilingual manuals; however, responses frequently diverged from frontline troubleshooting practice. This gap motivated a redesign toward a modular, human-authored retrieval pipeline with structured knowledge encoding and deterministic routing. The redesigned system comprised six modular components: (1) a hybrid rule-based and LLM-driven troubleshooting classifier for initial query routing; (2) a semantic intent and topic classifier using BGE-M3 embeddings and a k-nearest neighbors (k-NN) search to retrieve top-$k$ candidate intents/documents; (3) topic-disambiguation logic to resolve ambiguous or underspecified requests; (4) a slot-filling engine implemented via an LLM to extract required entities and parameters; (5) a procedure-selection component for retrieving relevant troubleshooting steps; and (6) a response-formatting module that renders outputs in a consistent user-facing structure (Fig.  \ref{fig:LLM_RAG_workflow}). 

A companion web-based knowledge management interface enabled support staff to update topics, specify slot constraints, log unresolved queries, and track coverage gaps, embedding end-user governance directly into system behavior.

System performance was evaluated using stakeholder-aligned metrics. Classification components were assessed with micro-averaged Precision, Recall, and F1-score, while ranking of classified topics were monitored using Hit Rate@k, MRR, and NDCG. These measures were tracked continuously and linked to organizational quality indicators such as procedural compliance and resolution accuracy.

Two comprehensive developer diaries were maintained across the full system lifecycle. The diaries were independently authored by two engineers directly involved in system design and implementation, drawing on shared project artifacts such as evaluation logs, deployment records, unresolved query trackers, and knowledge-base update logs. To mitigate author bias, each engineer cross-reviewed the other’s entries, followed by verification by a third researcher not involved in day-to-day development and validation by an external HCAI expert. The diaries documented five major pipeline stages: (1) stakeholder requirement gathering and alignment, (2) proof of concept and failure validation, (3) data curation and knowledge schema co-design, (4) iterative system redesign and performance management, and (5) integration, internal trial, and deployment preparation. To ensure the credibility and internal consistency of the diary entries, records were independently authored, mutually reviewed, and verified against contemporaneous project artifacts. Where inconsistencies or ambiguities in event descriptions arose, these were discussed and resolved collaboratively before inclusion in the analytic dataset.

\subsection{Data Source 2: AI Expert Interview Study}
We recruited eight participants through purposive, criterion-based sampling using our professional and research networks. Eligibility required at least three years of professional experience developing, deploying, evaluating, or governing enterprise AI applications. The final cohort included academic and industry participants, including data scientists, AI engineers, and product managers.

Data were collected using a semi-structured interview protocol, allowing for consistent thematic coverage across participants while permitting exploration of real-world experience and insights. The protocol contained thirteen open-ended questions (e.g., “How do you translate a business problem into an AI-solvable task, and who is involved in defining that problem?”). Core thematic areas included the translation of business requirements into technical specifications, mechanisms for ensuring data and output quality, optimal divisions of labor between human and AI agents, strategies for system monitoring and maintenance, approaches to interpretability, and governance structures for accountability and risk mitigation. To illustrate the type of practices discussed in relation to early-stage problem definition, one participant noted: “We have to conduct something similar to an interview … We have to ask them, create a requirement, and then confirm.” All participants provided informed consent, and interview responses were transcribed and prepared for analysis.

\subsection{Preparation and Analytical Approach}
\begin{figure}
    \centering
    \includegraphics[width=1\linewidth]{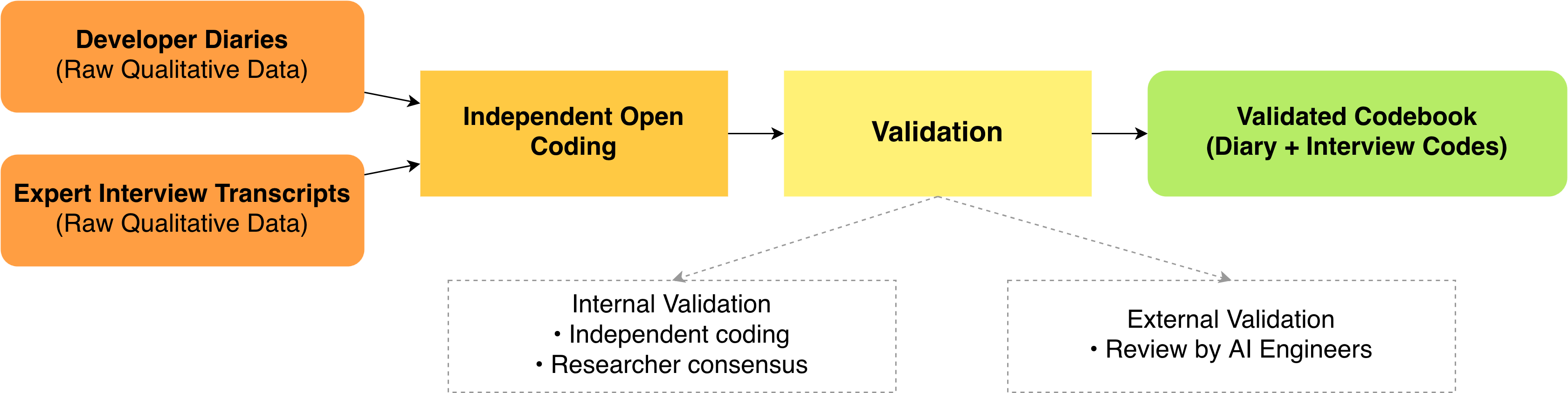}
    \caption{Open coding workflow}
    \label{fig:codebook_generation}
\end{figure}

The diaries and interview data served as the primary qualitative sources for open coding in this study. Open coding followed standard thematic analysis procedures: researchers independently reviewed diary entries and interview transcripts, segmented the data into meaningful units (phrases, sentences, or brief passages), and assigned descriptive codes capturing salient meanings, decisions, and practices. Both diary and interview datasets were coded using the same open coding protocol to ensure methodological consistency across sources (Fig. ~\ref{fig:codebook_generation}). To support internal validation at this stage, initial code assignments were cross-checked by the researchers and external AI engineers, and discrepancies in code interpretation or labeling were discussed and resolved through consensus. This process yielded approximately 2,300 initial codewords reflecting the breadth of empirical observations in the raw data.

\begin{figure}
    \centering
    \includegraphics[width=1\linewidth]{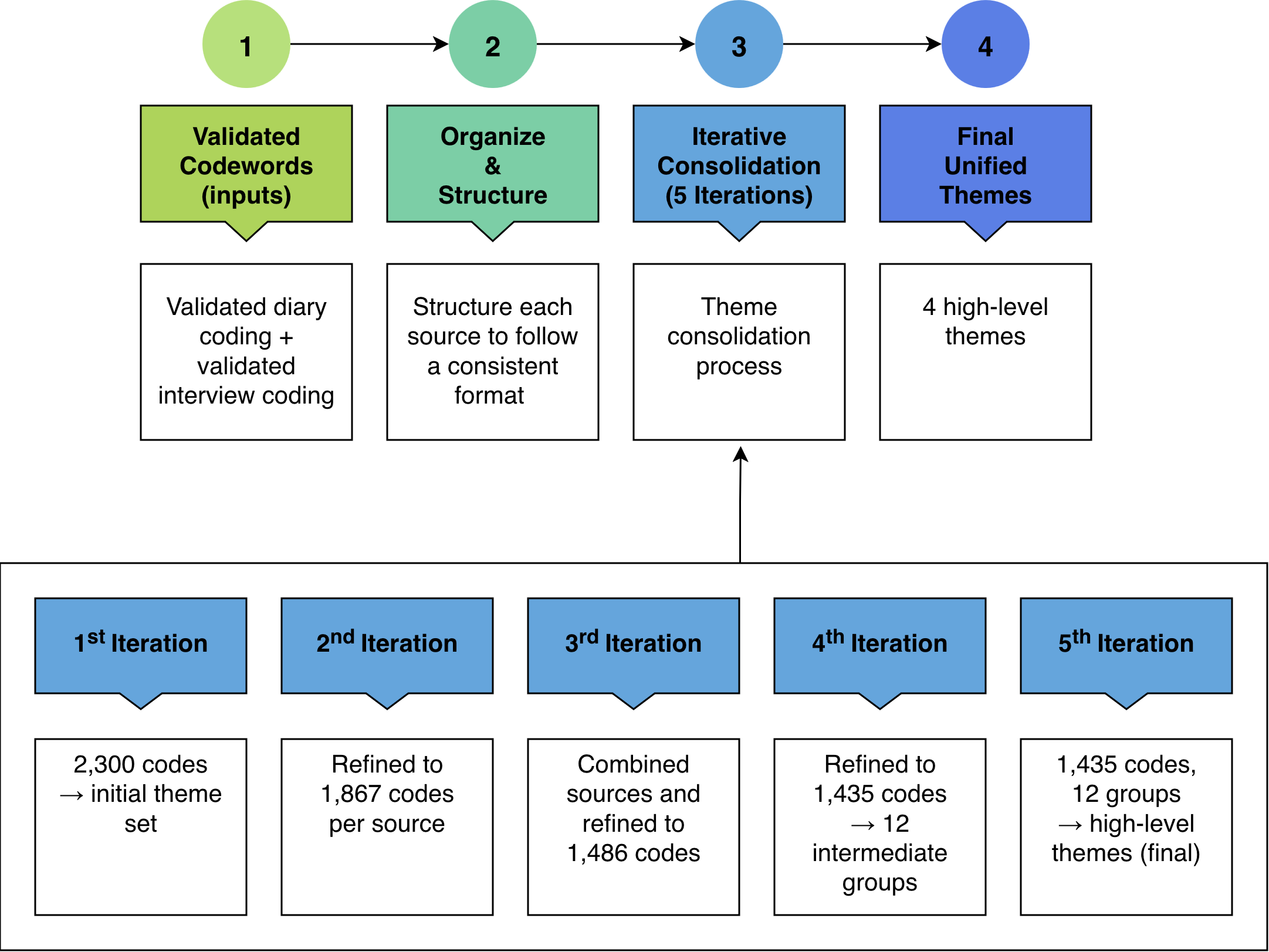}
    \caption{Thematic analysis process}
    \label{fig:thematic_analysis_cycles}
\end{figure}

Following open coding, an iterative thematic synthesis was conducted to consolidate and refine the codewords (Fig. ~\ref{fig:thematic_analysis_cycles}). Across four refinement iterations, duplicate and overlapping codes were progressively merged or reassigned. By the fourth iteration, code reduction reached its limit without loss of nuance or analytic specificity. A fifth iteration focused on abstraction, consolidating the remaining patterns into four high-level themes with coherence and relevance across the AI system lifecycle. The fifth iteration marked a stopping point, beyond which further iteration did not improve analytical resolution.

All materials were used with permission for academic research purposes. Interview participants provided informed consent, and sensitive data were stored in access-controlled repositories.


\section{RESULTS}

The thematic analysis yielded four high-level themes that capture recurring patterns in how human judgment, authority, and coordination manifested during AI system development and deployment. These themes represent analytically derived groupings grounded in diary entries and expert interviews, with each theme aggregating multiple empirically observed patterns that co-occurred across data sources. The themes are interdependent and reflect overlapping dimensions of governance, iteration, constraints, and collaboration, rather than a one-to-one mapping to individual research questions. The four themes and their constituent subthemes are presented below, along with representative empirical patterns and supporting evidence from diary observations and interview accounts. 

\subsection{Theme 1: AI Governance and Human Authority}
This theme captures how human authority, accountability, and oversight were enacted in practice, revealing governance as an emergent, situated process shaped by organizational context, uncertainty, risk considerations, and emerges through four interrelated subthemes, highlighted in bold. \textit{\textbf{Human Roles, Authority, and Accountability}} describes how decision authority was dynamically negotiated based on expertise, confidence in system outputs, and task-specific context, frequently extending beyond formally defined organizational roles. \textit{\textbf{Reliability, Risk, and Safety Oversight}} reflects the application of human judgment during data validation, model evaluation, and deployment monitoring, particularly in scenarios where automated criteria were insufficient to ensure reliable or safe operation. \textit{\textbf{Requirements, Stakeholders, and Governance Constraints}} highlights how system requirements evolved through ongoing negotiation among technical teams, product stakeholders, and compliance functions, rather than being fixed at initial design stages. Finally, \textit{\textbf{Strategic and Organizational Context}} emphasizes how broader business objectives and organizational priorities shaped model selection, deployment timing, and the overall scope of the AI system. Collectively, these subthemes show that HITL governance is realized through ongoing negotiation of authority, risk, and responsibility across organizational roles and development stages.

\subsection{Theme 2: Human-in-the-Loop Iterative Refinement}
This theme reflects the cyclical and feedback-driven nature of AI development, in which system understanding evolved through repeated experimentation and reassessment rather than linear progression, and is derived from a single subtheme, highlighted in bold. \textit{\textbf{Model Development and Experimentation}} describes how participants iteratively tested alternative modeling approaches, evaluated outcomes using both quantitative metrics and expert judgment, and revised system designs when automated performance signals conflicted with domain-informed assessment. Taken together, these patterns highlight iterative refinement as a HITL learning process in which progress depends on continuous interaction between automated signals and human judgment.

\subsection{Theme 3: AI System Lifecycle and Operational Constraints}
This theme captures how AI system development was shaped by practical constraints related to resources, timelines, and organizational structure, resulting in pragmatic trade-offs across design and operational decisions, and and emerges through five interrelated subthemes, highlighted in bold. \textit{\textbf{AI System Design, Architecture, and Integration}} examines how architectural choices were constrained by infrastructure availability, technical feasibility, and organizational context, often favoring incremental or modular solutions over idealized designs. \textit{\textbf{Cross-Functional Collaboration and Communication}} underscores the need for sustained coordination across technical and non-technical roles, as teams balanced competing priorities, evaluation criteria, and stakeholder expectations. \textit{\textbf{Data Management, Quality, and Validation}} addresses how human involvement guided data selection, quality assessment, and documentation practices, including explicit trade-offs between methodological rigor and development speed. \textit{\textbf{Deployment, Operations, and Infrastructure}} characterizes how participants monitored system behavior post-deployment and responded to operational issues under time, staffing, and support limitations. \textit{\textbf{Project, Lifecycle, and Resource Management}} situates these practices within broader constraints imposed by staffing capacity, compute budgets, and delivery deadlines, which shaped prioritization decisions, scope reduction, and the deferral of non-critical system improvements. Taken together, these subthemes show that HITL practices are fundamentally shaped by project-level and operational constraints, requiring teams to continuously negotiate trade-offs between technical rigor, organizational capacity, and delivery pressures.

\subsection{Theme 4: Human–AI Team Collaboration and Coordination}
This theme captures collaborative practices that enabled shared understanding, collective decision-making across roles and areas of expertise, and emerges through two interrelated subthemes, highlighted in bold. \textit{\textbf{Evaluation, Metrics, and Performance Assessment}} describes how evaluation decisions were negotiated collaboratively, particularly when quantitative performance metrics conflicted with operational constraints or product goals. \textit{\textbf{Interaction Design, Prompting, and Explainability}} details how participants employed interface design choices and prompting strategies to probe, interpret, and iteratively refine system behavior, thereby supporting shared understanding of system capabilities and limitations across team boundaries. Overall, this theme frames collaboration as an enabling condition for coordinated human oversight rather than a peripheral organizational activity. 

\begin{figure}
    \centering
    \includegraphics[width=1\linewidth]{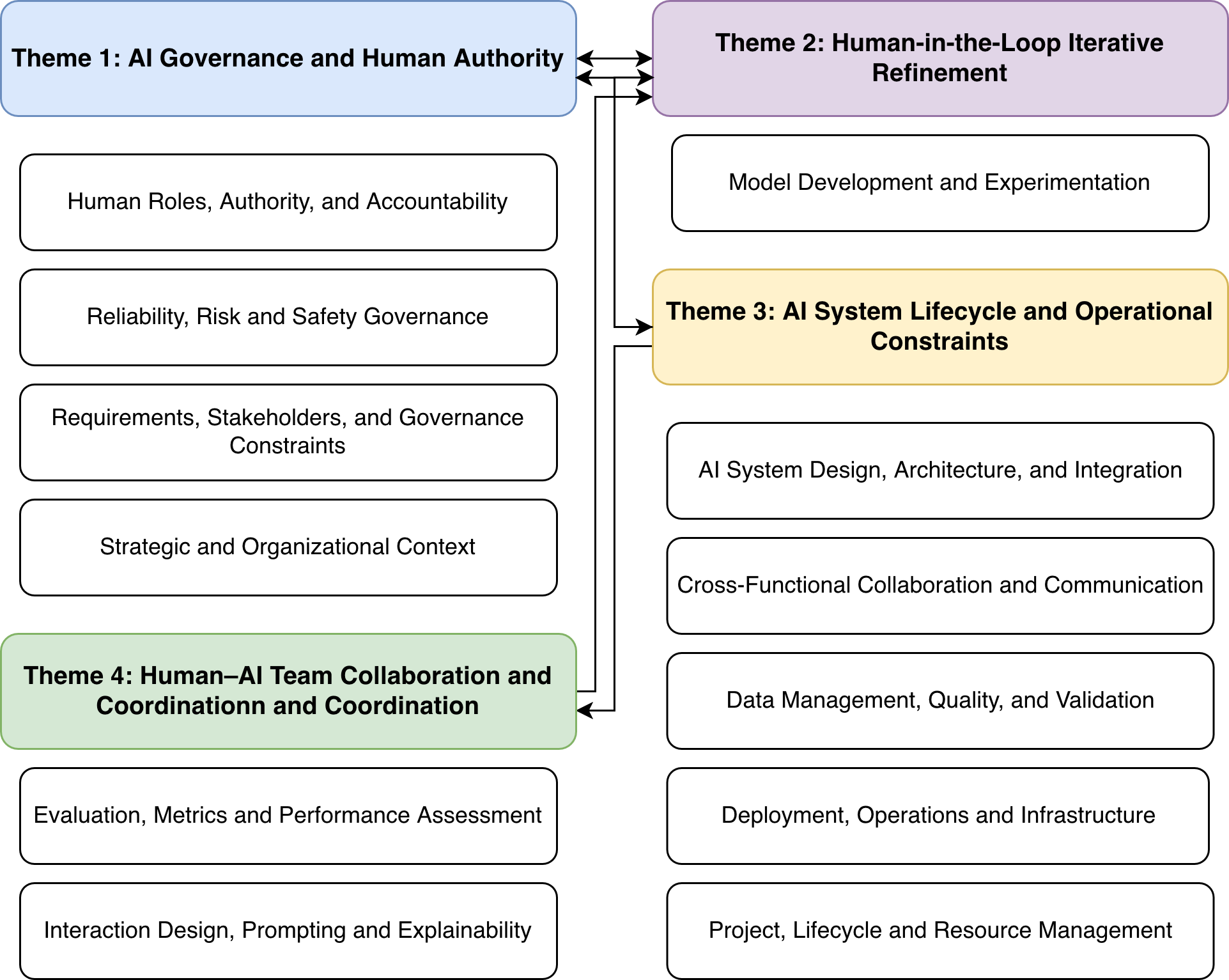}
    \caption{Hierarchical relationship of four high-level themes}
    \label{fig:theme_relationships}
\end{figure}

\section{Discussion and Future Work}
The four themes indicate that human oversight in AI application development is enacted less as a single checkpoint than as recurring organizational work distributed across the lifecycle. In particular, \textit{AI Governance and Human Authority} is visible in the negotiation of roles and accountability, risk and reliability oversight, and the iterative (re)definition of requirements under organizational priorities; \textit{Human-in-the-Loop Iterative Refinement} reflects experimentation guided by both quantitative signals and domain judgment; \textit{AI System Lifecycle and Operational Constraints} captures how architecture, data management, deployment operations, and project resources bound feasible oversight; and \textit{Human--AI Team Collaboration and Coordination} highlights how teams jointly negotiate evaluation practices and use interaction design, prompting, and explainability to align understanding across roles. Taken together, these results suggest that human agency in practice is situated, negotiated, and context-dependent, extending beyond purely procedural oversight accounts of HITL \cite{WU2022364, GRONSUND2020101614}.

Although analytically separable, the themes function as a coupled set of socio-technical dynamics. Governance decisions shape what is acceptable to test and deploy, iterative refinement surfaces mismatches that drive requirement renegotiation, and operational constraints delimit which oversight mechanisms can be sustained; collaboration practices mediate trade-offs when objectives and metrics conflict across technical, organizational, and regulatory priorities. Consistent with the diary-based lifecycle perspective, the salience of these dynamics shifts over time, with early work emphasizing experimentation and architectural design and later work foregrounding reliability oversight and operational constraints. This pattern aligns with prior characterizations of AI development as a socio-technical process embedded within organizational structures \cite{shneiderman2020humancenteredartificialintelligencereliable, Holstein_2019}.

A recurring tension emerged between automated evaluation metrics and human judgment. While automated metrics supported scalability, human evaluation was time-intensive and applied selectively, particularly when metrics diverged from operational assessments. This tension reflects fairness and accountability concerns identified in prior work \cite{10.1145/3287560.3287598}. Documentation further emerged as a key governance mechanism, preserving traceability, accountability, and organizational memory across the system lifecycle \cite{raji2020closingaiaccountabilitygap}.

Despite scope constraints, reliance on self-reported accounts, and a limited sample size in the AI expert interviews, this study offers empirically grounded insight into how human judgment, authority, and coordination are enacted across the AI development lifecycle. This paper contributes an empirical synthesis of HITL practices in AI application development, demonstrating how governance arrangements, iterative development dynamics, operational constraints, and cross-functional collaboration jointly shape human oversight across the system lifecycle. Grounded in diary-based evidence and expert interviews, the resulting themes provide a structured foundation for operational HITL guidance.

Future work will operationalize these findings into a structured, lifecycle-oriented HITL framework specifying role definitions, decision checkpoints, feedback protocols, and governance mechanisms aligned with established standards, and will validate the framework through application to an analytics-support chatbot performing business intelligence tasks. Future research should then examine the framework’s transferability to higher-risk domains (e.g., healthcare and public services) and investigate the organizational and technical infrastructures required to scale human oversight from project-level implementation to institutional practice.

\section{Conclusion}
This paper addresses a key gap in AI application development: the lack of structured, organization-centered operational guidance for implementing HITL principles. Through a multi-source qualitative study, we derived four validated themes that capture the core socio-technical dynamics required to operationalize HITL in practice. These findings shift attention from isolated human intervention points toward the organizational processes that sustain accountability and oversight in AI systems. Future work will formalize these insights into a prescriptive, lifecycle-oriented HITL framework and validate it across application domains.

\bibliographystyle{IEEEtran}
\bibliography{ref}

\vspace{12pt}

\end{document}